\documentclass[]{interact}
\usepackage{epstopdf}
\usepackage{subfigure}
\usepackage{caption}
\usepackage{mathptmx}
\usepackage{graphicx}
\usepackage{bm}
\usepackage{float}
\usepackage[graphicx,array,empheq,amsmath,amsthm,amssymb,amsfonts]{xcolor}
\usepackage[numbers,sort&compress]{natbib}
\bibpunct[, ]{[}{]}{,}{n}{,}{,}

\makeatletter
\def\NAT@def@citea{\def\@citea{\NAT@separator}}
\makeatother
\makeatletter
\def\NAT@def@citea{\def\@citea{\NAT@separator}}
\makeatother
\begin{document}
\articletype{Original Article}

\title{Solitary waves and double layers in complex plasma media}

\author{
\name{A A Mamun\textsuperscript{a,c}\thanks{\textsuperscript{c}Also at Wazed Miah Science Research Centre, Jahangirnagar University, Savar, Dhaka-1342, Bangladesh. Email: mamun$_{-}$phys@juniv.edu} and Abdul Mannan\textsuperscript{a,b}\thanks{Corresponding author: Abdul Mannan. Email: abdulmannan@juniv.edu; Telephone: +4915210286280; Fax: +49-345-55-27005}}
\affil{\textsuperscript{a}Department of Physics, Jahangirnagar University, Savar, Dhaka-1342, Bangladesh\\
\textsuperscript{b}Institut f\"{u}r Mathematik, Martin Luther Universit\"{a}t Halle-Wittenberg, D-06099 Halle, Germany}
}
\maketitle
\begin{abstract}
A complex plasma medium (containing Cairns nonthermal electron species, adiabatically warm inertial ion species, and stationary positively charged dust (PCD) species (making a plasma system very complex) is considered. The effects of PCD species, nonthermal electron species, and adiabatic ion-temperature on ion-acoustic (IA) solitary waves (SWs) and double layers (DLs) are investigated by the pseudo-potential approach, which is valid for the arbitrary amplitude time-independent SWs and DLs. It is observed that the presence of the PCD species reduces the phase speed of the IA waves, and consequently supports the IA subsonic compressive SWs in such electron-ion-PCD plasmas. On the other hand, the increase in both adiabatic ion-temperature and the number of nonthermal or fast electrons causes to reduce the possibility for the formation of the subsonic SWs, and thus convert subsonic SWs into supersonic ones. It is also observed that after at a certain value of the nonthermal parameter, the IA supersonic SWs with both positive and negative potentials as well as the DLs with only negative potential exist. The applications of the work in space environments (viz. Earth's mesosphere, cometary tails, Jupiter's magnetosphere, etc.) and laboratory devices, where the warm ion and nonthermal electron species along with PCD species have been observed, are briefly discussed.
\end{abstract}
\section{Introduction}
\label{Intro}
Nowadays, the existence of positively charged dust (PCD) species in electron-ion plasmas received a renewed interest because of their vital role in modifying existing features as well as introducing new features of linear and nonlinear ion-acoustic (IA) waves propagating in many space plasma environments [viz. Earth's mesosphere \cite{Havnes96,Gelinas98,Mendis04}, cometary tails \cite{Horanyi96,Mamun04}, Jupiter's surroundings \cite{Tsintikidis96}, Jupiter's magnetosphere \cite{Horanyi93}, etc.] and laboratory devices \cite{Khrapak01,Fortov03,Davletov18}, where in addition to electron-ion plasmas, the PCD species have been observed. Three principal mechanisms by which the dust species becomes positively charged \cite{Chow93,Rosenberg95,Rosenberg96,Fortov98} are as follows:
\begin{itemize}
\item{The photoemission of electrons from the dust grain surface induced by the flux of high energy photons \cite{Rosenberg96}.}
\item{The thermionic emission of electrons from the dust grain surface by the intense radiative or thermal heating \cite{Rosenberg95}.}
\item{The secondary emission of electrons from the dust grain surface by the impact of high energetic plasma particles like electrons or ions \cite{Chow93}}.
\end{itemize}
The dispersion relation for the IA waves in an electron-ion-PCD plasma system (containing inertialess isothermal electron species, inertial cold ion species, and stationary PCD species) is given by \cite{MS20}
\begin{eqnarray}
\frac{\omega}{kC_i}=\frac{1}{\sqrt{1+\mu+k^2\lambda_D^2}},
\label{IA-dispersion1}
\end{eqnarray}
where $\omega=2\pi f$ and $k=2\pi/\lambda$ in which $f$ ($\lambda$) is the IA  wave frequency (wavelength); $C_i=(z_ik_BT_e/m_i)^{1/2}$ is  the IA speed in which $k_B$ is the Boltzmann constant, $T_e$ is the electron temperature, and $m_i$ is the ion mass; $\lambda_D=(k_BT_e/4\pi z_in_{i0}e^2)^{1/2}$ is  the IA wave-length scale in which $n_{i0}$ ($z_i$) is the number density (charge state) of the ion species at equilibrium, and $e$ is the magnitude of the charge of an electron; $\mu=z_dn_{d0}/z_in_{i0}$ with $n_{d0}$ ($z_d$) being the number density (charge state) of the PCD species at equilibrium. This means that $\mu=0$ corresponds to the usual electron-ion plasma, and $\mu\rightarrow\infty$ corresponds to electron-dust plasma \cite{Mamun04,Khrapak01,Fortov03,Davletov18}. Thus, $0<\mu<\infty$ is valid for the electron-ion-PCD plasmas. The dispersion relation defined by
(\ref{IA-dispersion1}) for the long-wavelength limit (viz. $\lambda\gg\lambda_D$) becomes
\begin{eqnarray}
\frac{\omega}{kC_i}\simeq\sqrt{\frac{1}{1+\mu}}.
\label{IA-dispersion2}
\end{eqnarray}
The dispersion relation (\ref{IA-dispersion2}) indicates that the phase speed decreases with the rise of the value of $\mu$. This new feature of the IA waves (continuous as well as periodic compression and rarefaction or vise-versa of the positive ion fluid) is introduced due to the reduction of the space charge electric field  by the presence of PCD.

Recently, based on this new linear feature, Mamun and Sharmin \cite{MS20} and Mamun \cite{Mamun21} have shown the existence of subsonic shock and SWs, respectively, by considering the assumption of Maxwellian electron species and cold ion species. The IA waves in different plasma systems composed of ions and electrons have also been studied by a number of authors \cite{Zedan20,El-Monier20,Mehdipoor20}. However, the reduction of the IA wave phase speed by the presence of PCD species can also make the IA phase speed comparable with the ion thermal speed  $V_{Ti}=(k_BT_i/m_i)^{1/2}$ (where $T_i$ is the ion-fluid temperature) so that the effect of the ion-thermal pressure cannot be neglected. On the other hand, the electron species in space environments mentioned does not always follow the Maxwellian electron velocity distribution function. This means that the linear dispersion relation (\ref{IA-dispersion2}), and the works of Mamun and Sharmin \cite{MS20} and Mamun \cite{Mamun21} are valid for a cold ion fluid ($T_i=0$) limit and for the Maxwell electron velocity distribution function, which can be expressed in one dimensional (1D) normalized [normalized by $n_{e0}/V_{Te}$, where $V_{Te}=(k_BT_e/m_e)^{1/2}$ is the thermal speed of the electron species, and $v$ is normalized by $V_{Te}$] form as
\begin{equation}
f(v)= \frac{1}{\sqrt{2\pi}}\exp\left[-\frac{1}{2}(v^2-2\phi)\right],
\label{MDf}
\end{equation}
where $\phi$ is the IA wave potential normalized by  $k_BT_e/e$.

\begin{figure*}[htb]
\centering
\begin{tabular}{@{}cc@{}}
\includegraphics[width=0.55\textwidth]{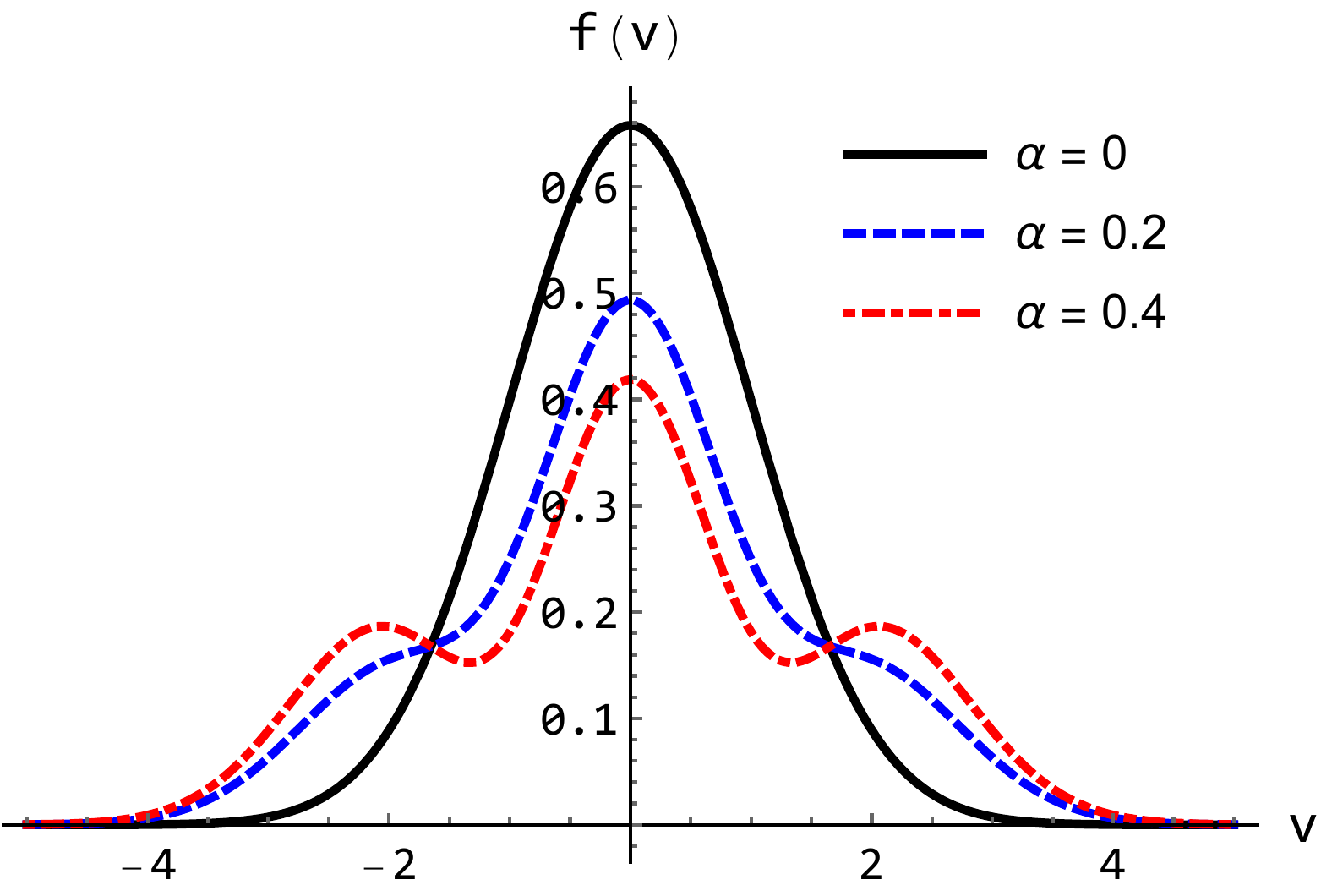} &
\includegraphics[width=0.42\textwidth]{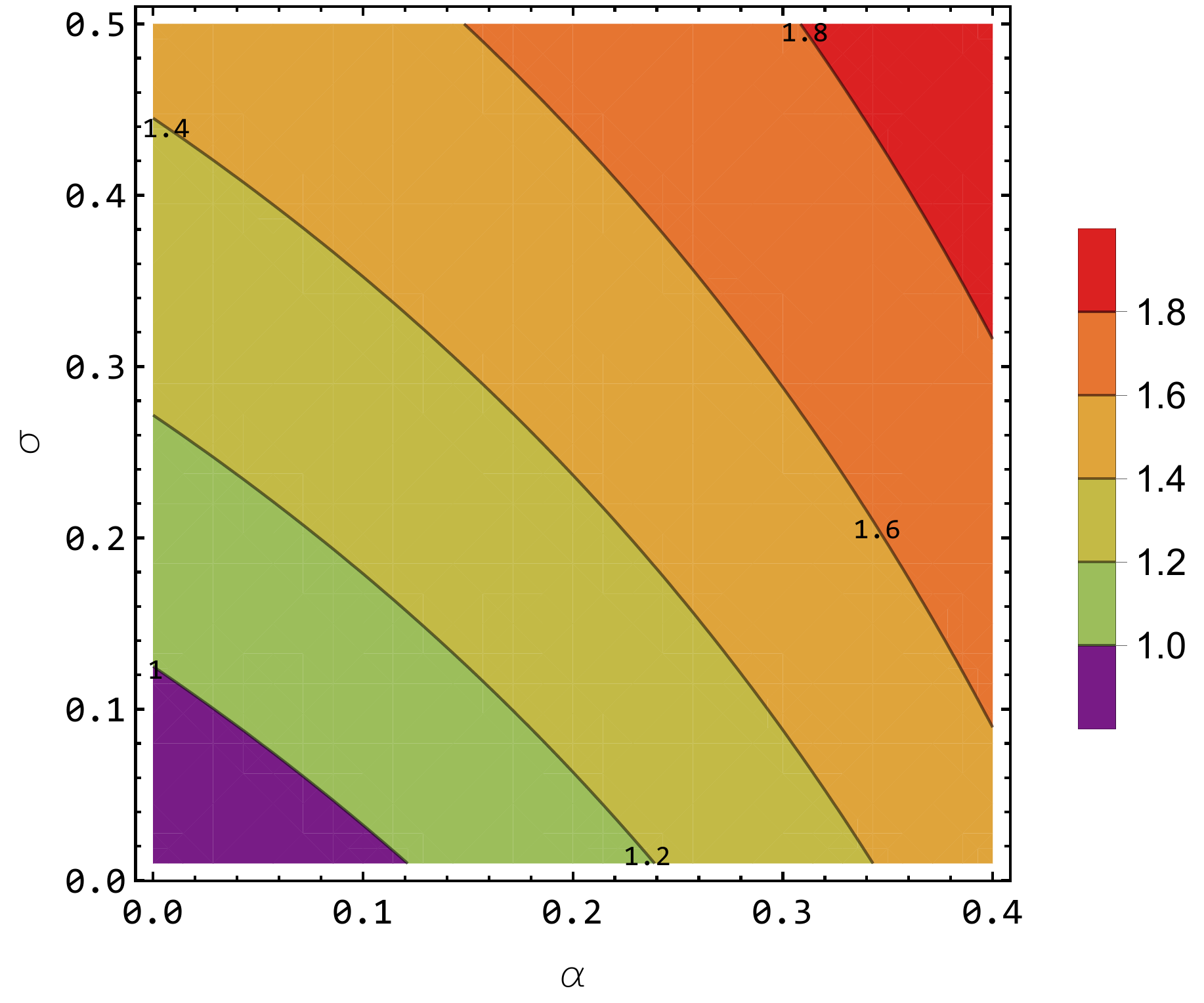}
\end{tabular}
\caption{The left panel shows the  curves representing the Cairns nonthermal  velocity distribution function  [defined by  equation (\ref{CDf})] for $\phi=0.5$ and different values of $\alpha$, whereas the right panel shows how the normalized phase speed ($\omega/kC_i$) of the IA waves [defined by equation (\ref{IA-dispersion3})] varies with $\sigma$ and $\alpha$ for $\mu=0.6$.}
\label{Fig1}
\end{figure*}

To overcome these two limitations, we consider (i) adiabatically warm ion fluid and (ii) nonthermal electron species following Cairns velocity distribution function, which can be similarly expressed in 1D normalized form as \cite{Cairns95}
\begin{equation}
f(v)= \frac{1+\alpha(v^2-2\phi)^2}{(1+3\alpha)\sqrt{2\pi}}\exp\left[-\frac{1}{2}(v^2-2\phi)\right],
\label{CDf}
\end{equation}
where $\alpha$ is a parameter determining the population of fast (energetic) particles present in the plasma system under consideration. We note that  equation (\ref{CDf}) is identical to equation (\ref{MDf}) for $\alpha=0$. Thus, how the nonthermal parameter $\alpha$ modifies the Maxwell distribution of the electron species is shown mathematically by equation (\ref{CDf}) and graphically by the left panel of figure \ref{Fig1}. On the other hand, including the effects of the Cairns nonthermal electron distribution ($\alpha$) and the adiabatic ion-temperature ($\sigma$), the dispersion relation for the long wavelength IA waves can be expressed as
\begin{eqnarray}
\frac{\omega}{kC_i}=\sqrt{\frac{1+3\alpha}{(1+\mu)(1-\alpha)}+3\sigma},
\label{IA-dispersion3}
\end{eqnarray}
where $\sigma=T_{i0}/z_iT_e$ with $T_{i0}$ is the ion-temperature at equilibrium. The dispersion relation
(\ref{IA-dispersion3}) indicates that as $\alpha$ and $\sigma$ increase, the phase speed of the IA waves increases. The is due to the enhancement of the space charge electric field by nonthermal electron species
and of the flexibility of the ion fluid by its temperature. The variation of the phase speed of the IA waves [defined by equation (\ref{IA-dispersion3})] with $\alpha$ and $\sigma$ is shown in the right panel of figure \ref{Fig1}.

The aim of this work is to investigate the combined effects of positively charged stationary dust species, Cairns nonthermal  electron distribution and adiabatic ion-temperature on the basic features of the IA solitary waves (SWs) and double layers (DLs) in electron-ion-PCD plasma system by the pseudo-potential approach \cite{Cairns95,Mamun97,Bernstein57}.

The manuscript is structured as follows. The equations describing the nonlinear dynamics of the IA waves in an electron-ion-PCD plasma are provided in section \ref{GE}. The combined effects of stationary PCD species, adiabatic ion-temperature and  nonthermally distributed electron species on IA SWs and DLs are investigated by the pseudo-potential approach in section \ref{SWs-DLs}. A brief discussion is finally presented in section \ref{DIS}.

\section{Governing equations}
\label{GE}
To investigate the nonlinear propagation of the IA waves defined by the equation (\ref{IA-dispersion3}), we consider an electron-ion-PCD plasma medium. The nonlinear dynamics of the IA waves propagating in such
an electron-ion-PCD plasma medium is described by
\begin{eqnarray}
&&\frac{\partial n_i}{\partial t}
+\frac{\partial}{\partial x} (n_iu_i) = 0,
\label{IA-b1}\\
&&\frac{\partial u_i}{\partial t} + u_i \frac{\partial u_i}{\partial x}
=-\frac{\partial \phi} {\partial x}-\frac{\sigma}{n_i} \frac{\partial P_i} {\partial x},
\label{IA-b2}\\
&&\frac{\partial P_i}{\partial t}+ u_i\frac{\partial P_i}{\partial\xi}+\gamma P_i\frac{\partial u_i}{\partial x}=0,
\label{IA-b3}\\
&&\frac{\partial^2\phi}{\partial x^2}  =(1+\mu)n_e-n_i-\mu,
\label{IA-b4}
\end{eqnarray}
where $n_i$ is the ion number density normalized by $n_{i0}$;  $u_i$ is the ion fluid speed normalized by $C_i$; $P_i$ is the adiabatic  ion-thermal pressure normalized by $n_{i0}k_BT_{i0}$; $\gamma\,[=(2 +{\cal N})/{\cal N}]$ is the ion fluid adiabatic index with ${\cal N}$ being the number of degrees of freedom, which has the
value $1$ ($3$) for the 1D (3D) case so that in our present work ${\cal N}=1$ and $\gamma=3$; $t$ ($x$) is the time (space) variable normalized by $\omega_{pi}^{-1}$ ($\lambda_D$);
$n_e$ is the nonthermal electron number density normalized by $n_{e0}$, and is determined by integrating equation (\ref{CDf}) with respect to $v$ from $-\infty$ to $+\infty$, i.e. $n_e$ can be expressed as \cite{Mamun97}
\begin{eqnarray}
n_e=(1-\beta\phi+\beta\phi^2)\exp(\phi),
\label{ne}
\end{eqnarray}
with $\beta=4\alpha/(1+3\alpha)$. We note that for isothermal electron species $\gamma=1$, $T_i=T_{i0}$ and $P_i=n_ik_BT_i$, equations (\ref{IA-b1}) and (\ref{IA-b3}) are identical.
\section{SWs and DLs}
\label{SWs-DLs}
To study arbitrary amplitude IA SWs and DLs, we employ the
pseudo-potential  approach \cite{Cairns95,Mamun97,Bernstein57} by assuming
that all dependent variables in equations (\ref{IA-b1})--(\ref{IA-b4}) depend only on a
single variable  $\xi= x - {\cal M}t$, where ${\cal M}$  is the Mach number (defined by $\omega/kC_i$). This transformation ($\xi=x-{\cal M}t$) along with the substitution of equation (\ref{ne}) into equation (\ref{IA-b4}) and $\gamma=3$ into equation (\ref{IA-b3}) as well as the use
of the steady state condition allow us to write
(\ref{IA-b1})--(\ref{IA-b4}) as
\begin{eqnarray}
&&{\cal M}\frac{d n_i}{d\xi}- \frac{d}{d\xi}(n_i u_i) =0,
\label{B1}\\
&&{\cal M}\frac{d u_i}{dl\xi}- u_i \frac{d u_i}{d\xi}=\frac{d\phi}{d\xi} +\frac{\sigma}{n_i}\frac{dP_i}{d\xi},
\label{B2}\\
&&{\cal M}\frac{d P_i}{d\xi}- u_i\frac{dP_i}{d\xi}-3P_i\frac{d u_i}{d\xi}=0,
\label{B3}\\
&&\frac{d^2\phi}{d\xi^2}  =(1+\mu)\left(1-\beta\phi+\beta\phi^2\right)\exp(\phi)-n_i-\mu.
\label{B4}
\end{eqnarray}
The appropriate conditions  (viz.  $n_i\rightarrow 1$ and $u_i\rightarrow 0$ at $\xi \rightarrow \pm \infty$) reduce (\ref{B1}) to
\begin{eqnarray}
&&u_i={\cal M}\left(1-\frac{1}{n_i}\right),
\label{B5}\\
&&n_i=\frac{{\cal M}}{{\cal M}-u_i}.
\label{B6}
\end{eqnarray}
The substitution of (\ref{B5}) into (\ref{B3}) gives rise to
\begin{eqnarray}
\frac{1}{n_i}\frac{dP_i}{d\xi}+3P_i\frac{d}{d\xi}\left(\frac{1}{n_i}\right)=0,
\label{B7}
\end{eqnarray}
which finally reduces to
\begin{eqnarray}
P_i=n_i^3,
\label{B8}
\end{eqnarray}
where the integration constant is found to be $1$ under the conditions that $P_i\rightarrow 1$ and $n_i\rightarrow 1$ at $\xi \rightarrow \pm \infty$.  Similarly, the substitution of (\ref{B5}) into equation (\ref{B2}) yields
\begin{eqnarray}
{\cal M}\frac{du_i}{d\xi} -u _i\frac{du_i}{d\xi}
- {\sigma} \frac{dP_i}{d\xi} +  \frac{\sigma}{{\cal M}} u_i
\frac{dP_i}{d\xi}=\frac{d\phi}{d\xi}.
\label{B9}
\end{eqnarray}
Again,  multiplying (\ref{B3}) by $\sigma/{\cal M}$ one can write
\begin{eqnarray}
\sigma\frac{dP_i}{d\xi} -\frac{\sigma}{{\cal M}}u_i \frac{dP_i}{d\xi}
-3P_i\frac{\sigma}{{\cal M}}\frac{du_i}{d\xi} = 0.
\label{B10}
\end{eqnarray}
Now, performing (\ref{B10})$-2\times$(\ref{B9}) we obtain
\begin{eqnarray}
3\sigma (P_i-1) -\frac{3\sigma}{{\cal M}}(P_iu_i)-2{\cal M}u_i+u_i^2 + 2\phi=0,
\label{B11}
\end{eqnarray}
where the integration constant is found to be $3\sigma$ under the conditions that $P_i\rightarrow 1$,  $n_i\rightarrow 1$, $u_i\rightarrow 0$, and $\phi\rightarrow 0$ at $\xi \rightarrow \pm \infty$. The substitution of equations (\ref{B5}) and (\ref{B8}) into equation (\ref{B11}) yields
 \begin{eqnarray}
 3\sigma n_i^4 -({\cal M}^2+3\sigma-2\phi)n_i^2+{\cal M}^2=0.
\label{B12}
\end{eqnarray}
This is the quadratic equation for $n_i^2$. Thus, the expression for $n_i$ can be expressed as
\begin{eqnarray}
n_i=\frac{1}{\sqrt{6\sigma}}\left[\sqrt{
\Psi-\sqrt{\Psi^2-12\sigma {\cal M}^2}}\right],
\label{ni}
\end{eqnarray}
where $\Psi={\cal M}^2 +3\sigma-2\phi$.
Now, substituting equation (\ref{ni}) into equation (\ref{B4}), we obtain
\begin{eqnarray}
\frac{d^2\phi}{d\xi^2}  =(1+\mu)\left(1-\beta\phi+\beta\phi^2\right)\exp(\phi)-\frac{1}{\sqrt{6\sigma}}\left[\sqrt{
\Psi-\sqrt{\Psi^2-12\sigma {\cal M}^2}}\right]-\mu,
\label{Poisson}
\end{eqnarray}
We finally multiply both side of equation (\ref{Poisson}) by ($d\phi/d\xi$) and integrating the resulting equation with respect to $\phi$, we obtain
\begin{eqnarray}
\frac{1}{2}\left(\frac{d\phi}{d\xi}\right)^2 +V(\phi,\mathcal{M}) =0,
\label{EI}
\end{eqnarray}
which represents an energy integral of a pseudo-particle of unit mass, pseudo time $\xi$, pseudo-position $\phi$ and pseudo-potential $V(\phi,\mathcal{M})$ is defined by
\begin{eqnarray}
&&V(\phi,\mathcal{M})=C_0+\mu\phi-(1+\mu)\left[1+\frac{4\alpha}{1+3\alpha}\left(3-3\phi+\phi^2\right)\right] \exp[\phi]\nonumber\\
&&\hspace*{2cm}-\frac{\sqrt{2}}{3\sqrt{3\sigma}}\left(\sqrt{\Psi-\sqrt{\Psi^2-12\sigma{\cal M}^2}}\right)\left(\Psi+\frac{1}{2}\sqrt{\Psi^2-12\sigma{\cal M}^2}\right),
\label{PP}
\end{eqnarray}
where
\begin{eqnarray}
C_0=(1+\mu)\left[1+\frac{12\alpha}{1+3\alpha}\right]+\sigma+{\cal M}^2
\label{C0}
\end{eqnarray}
is the integration constant, and it is chosen in such a way that $V(\phi,{\cal M})=0$ at $\phi=0$.

\begin{figure}[h!]
  \centering
  \subfigure[]{\includegraphics[width=0.48\textwidth]{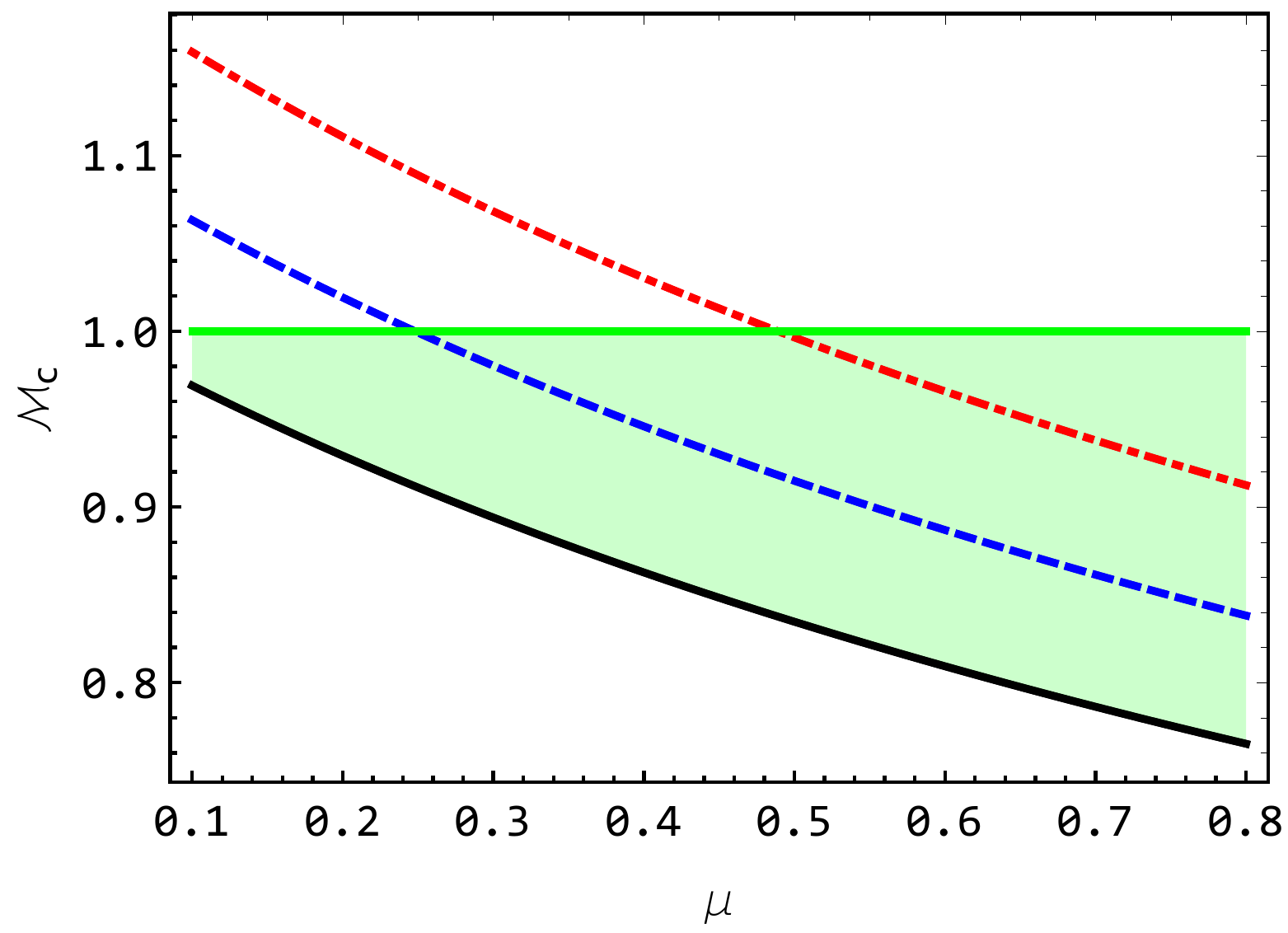}}
  \subfigure[]{\includegraphics[width=0.48\textwidth]{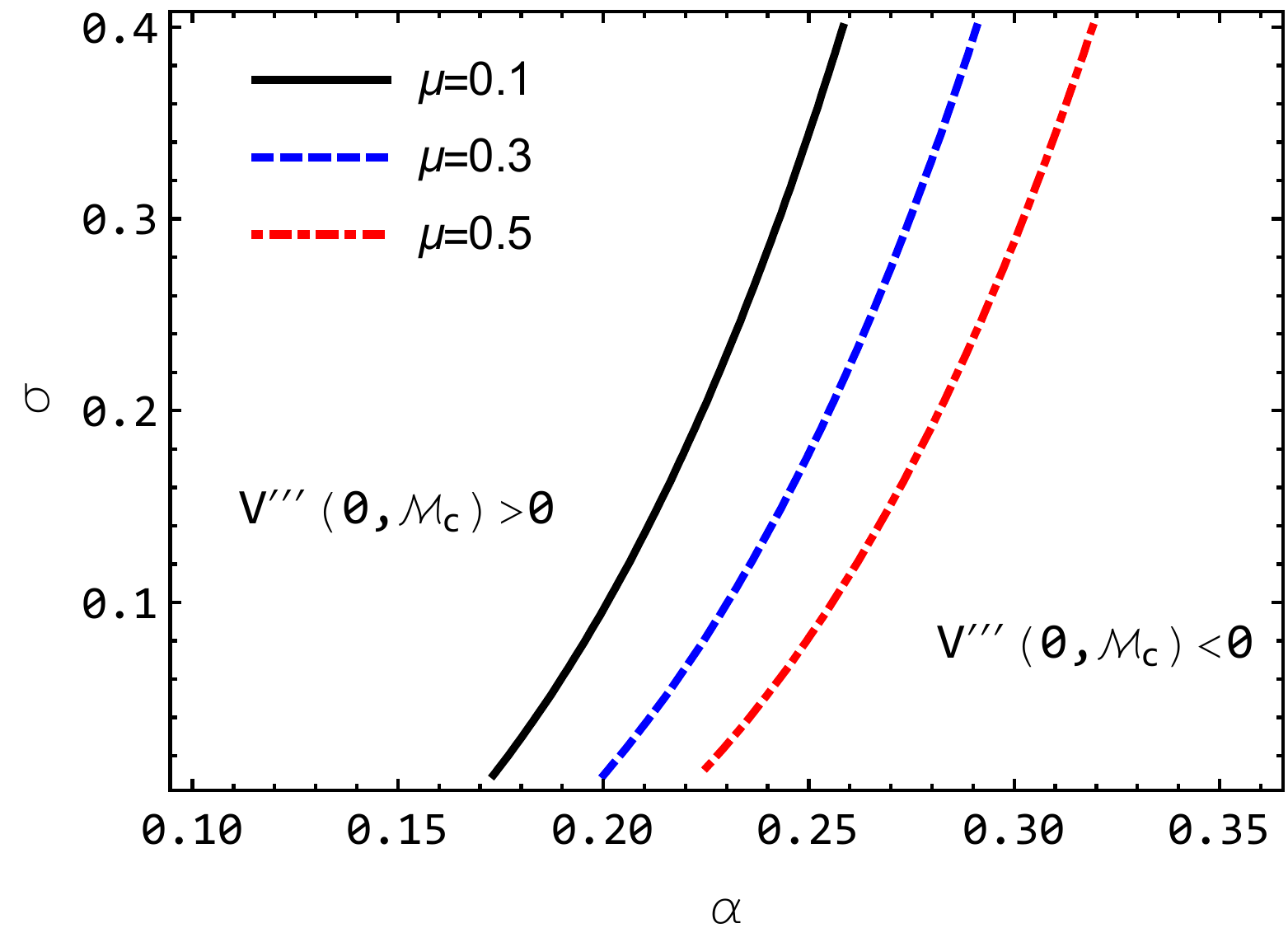}}
  \caption{(a) The variation of $\mathcal{M}_c$ with $\mu$ for $\sigma=0.01$ and $\alpha =0$ (solid curve),
 $\alpha = 0.05$ (dashed curve) and $\alpha = 0.1$ (dot-dashed curve). The shaded area corresponds to the existence of subsonic SWs; (b) The contour plot of $V'''(0,\mathcal{M}_c) = 0$ as a function of $\alpha$ and $\sigma$ for different values of $\mu$, viz $\mu=0.1$ (solid curve), $\mu=0.3$ (dashed curve) and $\mu=0.5$ (dot-dashed curve).}
  \label{Fig2}
\end{figure}

It is clear that $V(0,{\cal M}) =0$ is satisfied because of our choice of the integration constant, and $V'(0, {\cal M})=0$ is satisfied because of the equilibrium charge neutrality condition, where the prime denotes the derivative of $V(\phi, {\cal M})$ with respect to $\phi$.  So, the conditions for the existence of SWs and DLs are: (i) $V''(0,{\cal M})<0$ so that the fixed point at the origin is unstable (i.e. the convexity condition at the origin); (ii) $V'(\phi_m,{\cal M})>0$ for the SWs with $\phi>0$; (iii) $V'(\phi_m,{\cal M})< 0$ for the SWs with $\phi<0$; (iv) $V'(\phi_m,\mathcal{M}) = 0$ for the DLs, where $\phi_m$ is the amplitude of the SWs or DLs.
Thus, SWs or DLs exist if and only if $V''(0, {\cal M})<0$,  i.e. ${\cal M}>{\cal M}_c$, where
\begin{equation}
{\cal M}_c =\sqrt{ \frac{1+3\alpha}{(1+\mu)(1-\alpha)}+3\sigma}.
\label{Mc}
\end{equation}
We note that the expression for ${\cal M}_c$ [given by equation (\ref{Mc})] is identical to equation (\ref{IA-dispersion3}). The phase speed of the IA waves decreases and the possibility for the formation of subsonic IA SWs increases as the number of PCD species increases. This is depicted
in figure \ref{Fig1}(a).  On the other hand, the possibility for the formation of subsonic (supersonic) IA SWs decreases (increases) with the increase of the values of $\alpha$ and $\sigma$. This is shown in figure \ref{Fig1}(b). The ranges of the value of ${\cal M}$, viz. $\mathcal{M}_c < \mathcal{M} <1$ and $\mathcal{M}>\mathcal{M}_c>1$ determine the formation of subsonic and supersonic IA SWs, respectively.
The variation of $\mathcal{M}_c$ with $\mu$ and $\alpha$ for the fixed value of $\sigma$ is graphically shown in figure \ref{Fig2}(a), where the shaded (non-shaded) area represents the domain for the existence of subsonic (supersonic) SWs.

It is well known  \cite{Cairns95,Mamun97} that the sign of
\begin{equation}
  V'''(0,\mathcal{M}_c) = \frac{3(1-\alpha)^2(1+\mu)^2[1+3\alpha+4(1-\alpha)(1+\mu)\sigma]}{(1+3\alpha)^3}-(1+\mu)
 \label{polarity}
\end{equation}
determines either the existence of the IA SWs with $\phi>0$ or the coexistence of the IA SWs with $\phi>0$ and $\phi<0$. Thus, the IA SWs with $\phi>0$ [$\phi<0$ and $\phi>0$] will exist (coexist) if $V'''(0,\mathcal{M}_c)>0$ [$V'''(0,\mathcal{M}_c)<0$].
\begin{figure}[h!]
  \centering
  \subfigure[]{\includegraphics[width=0.48\textwidth]{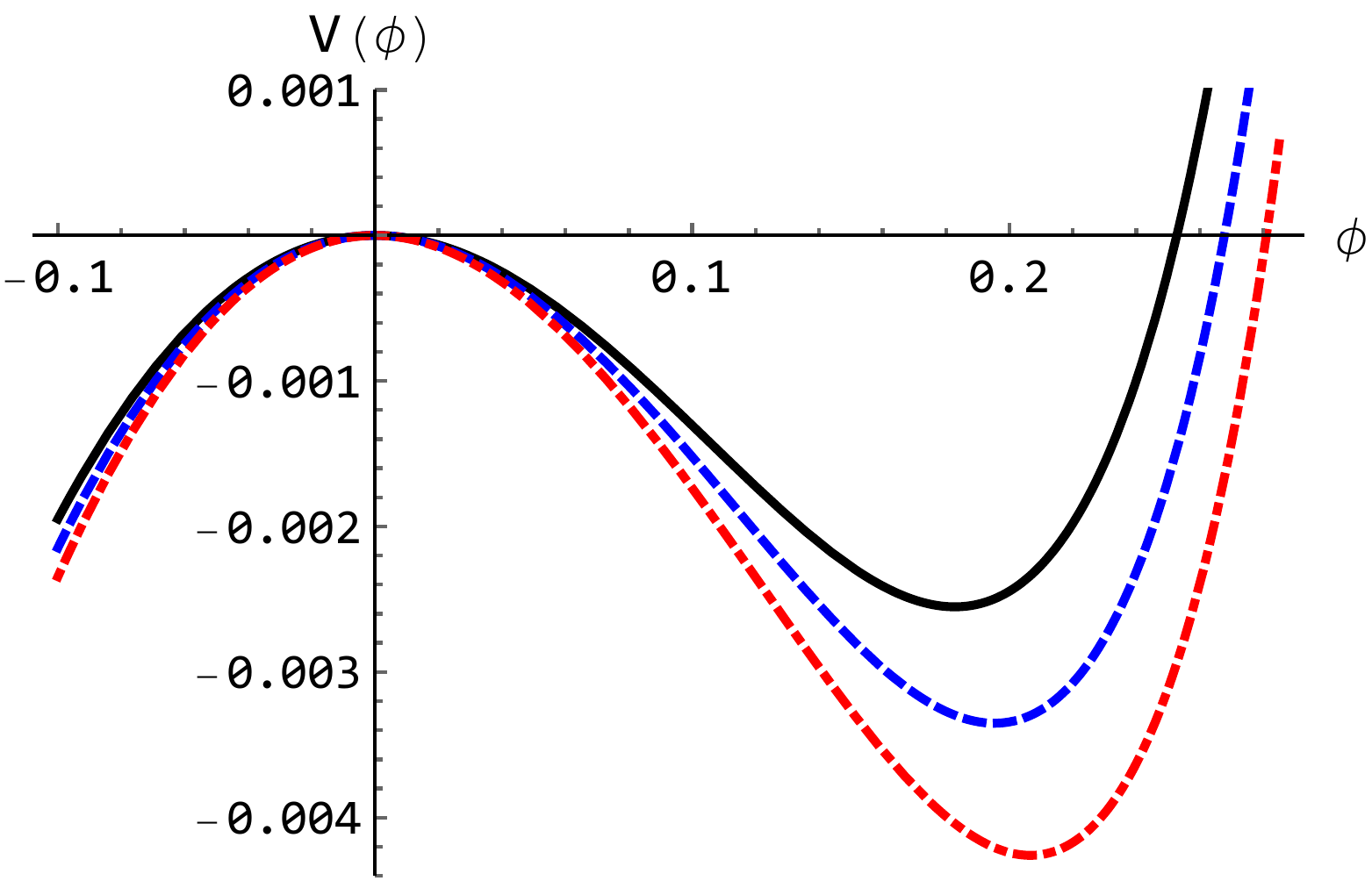}}
  \subfigure[]{\includegraphics[width=0.48\textwidth]{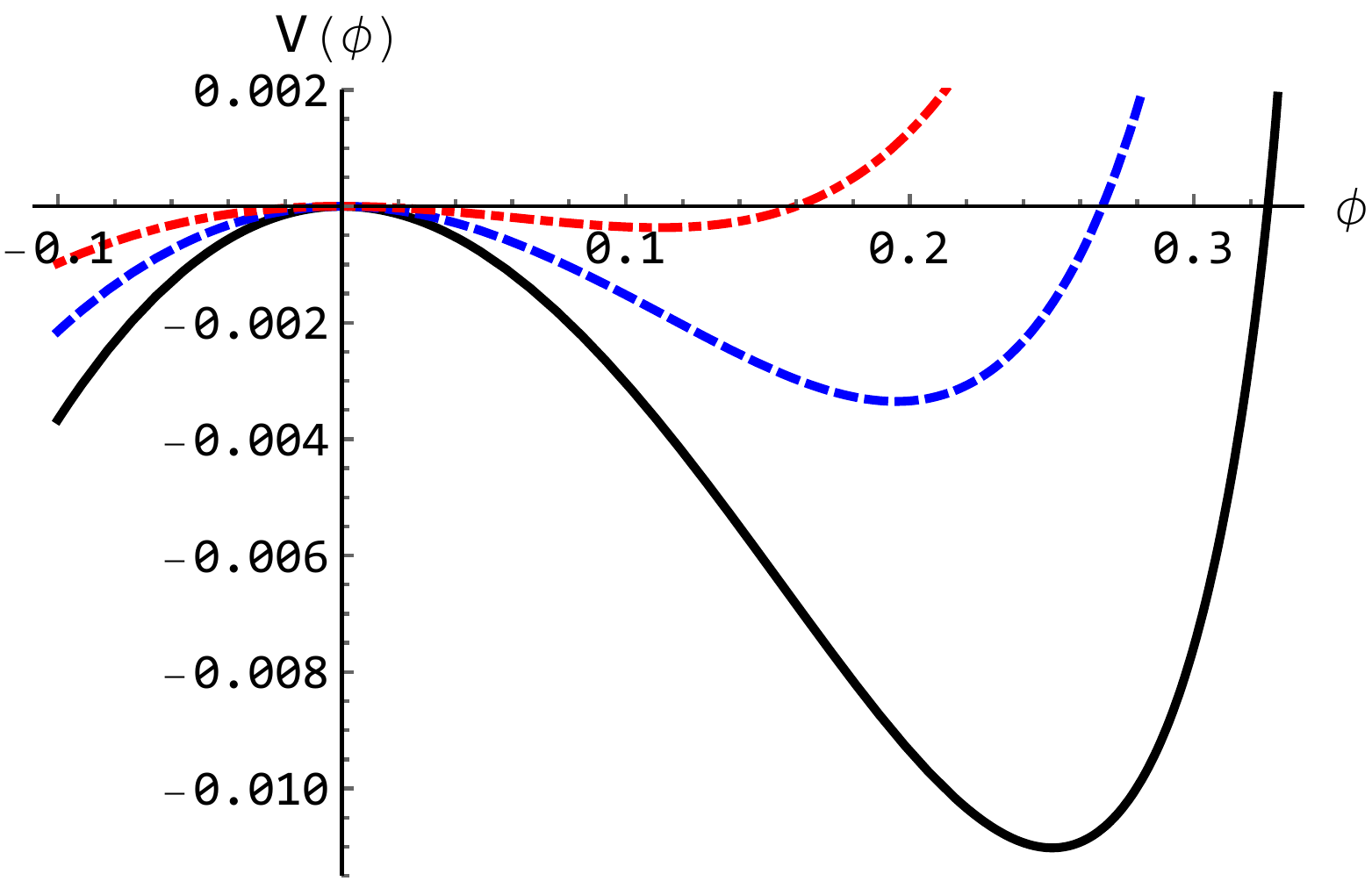}}
  \caption{The formation of the potential wells representing the subsonic  SWs
  (a) for $\alpha =0.05$, $\mu =0.7$ (solid curve), $\mu =0.75$ (dashed curve) $\mu =0.8$ (dot-dashed curve);
  (b) for $\mu =0.75$, $\alpha=0$ (solid curve), $\alpha =0.05$ (dashed curve) $\alpha =0.1$ (dot-dashed curve).
  The other parameters, which are kept fixed,  are ${\cal M}=0.985$ and $\sigma=0.01$.}
  \label{Fig3}
\end{figure}
\begin{figure}[h!]
    \centering
    \subfigure[]{\includegraphics[width=0.48\textwidth]{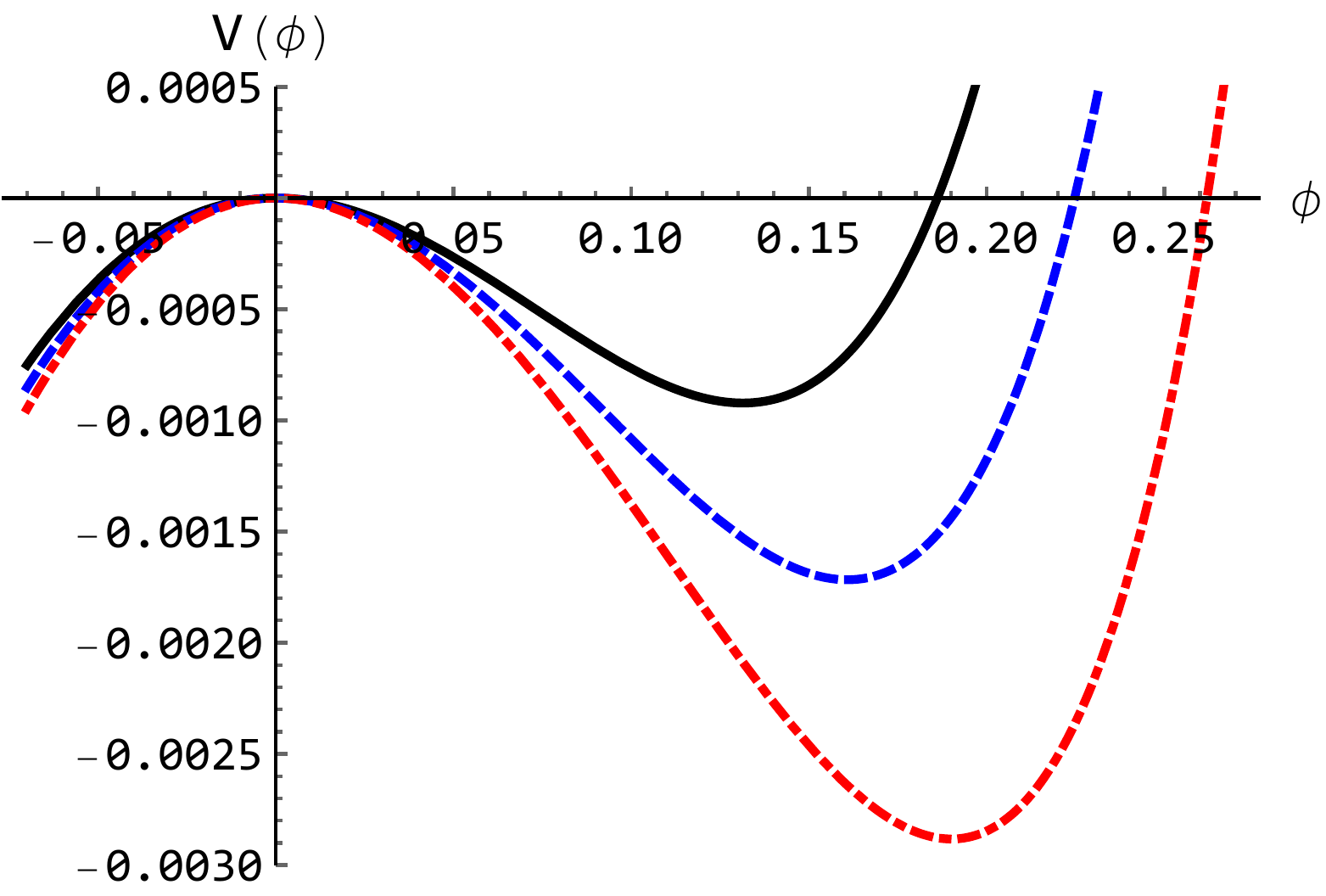}}
    \subfigure[]{\includegraphics[width=0.48\textwidth]{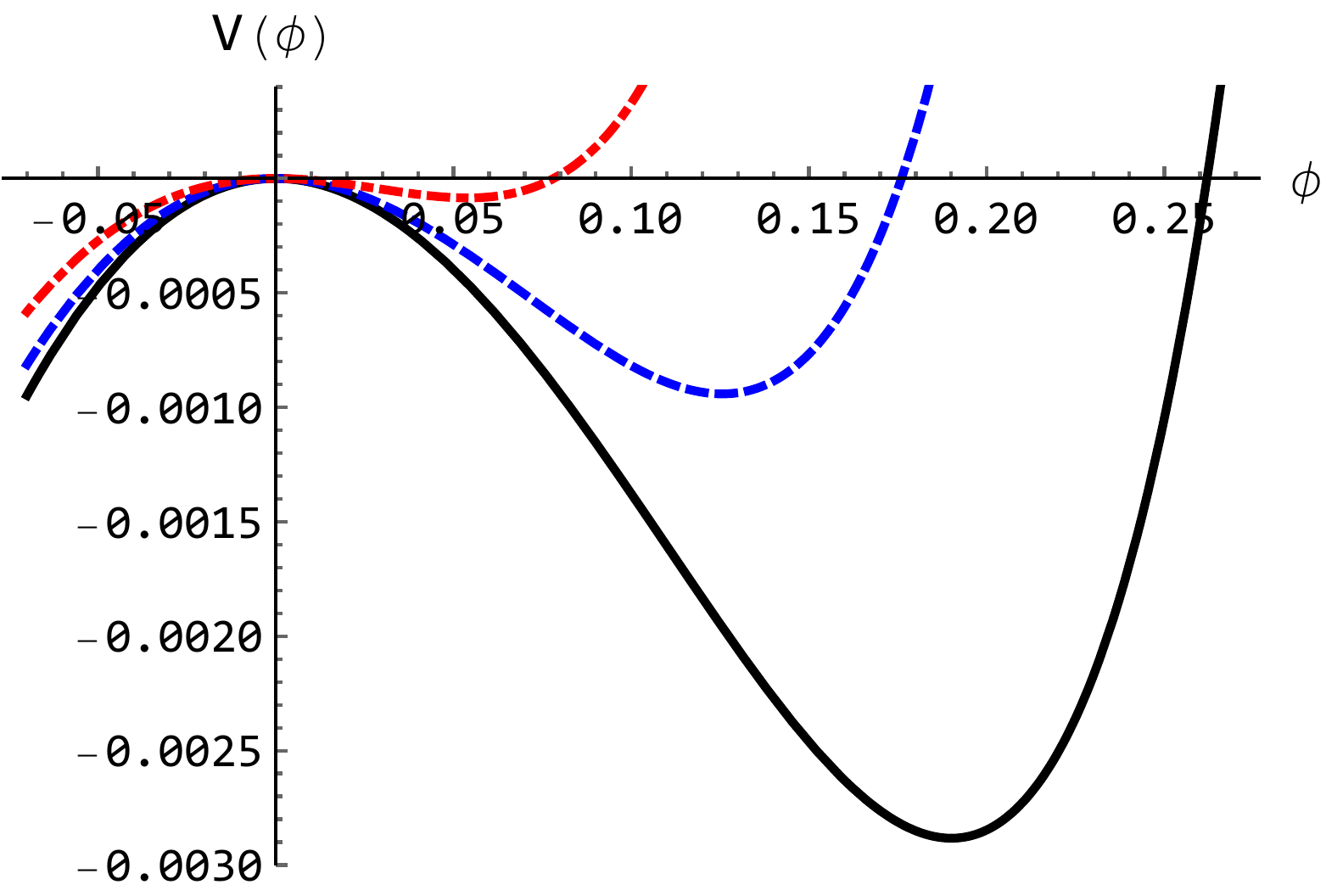}}
   \caption{The formation of the potential wells representing the subsonic  SWs
  (a) for $\sigma =0.01$, ${\cal M} =0.95$ (solid curve), ${\cal M}=0.97$ (dashed curve), ${\cal M}=0.99$ (dot-dashed curve); (b) for ${\cal M}=0.99$, $\sigma=0.01$ (solid curve), $\sigma =0.03$ (dashed curve) $\sigma =0.06$ (dot-dashed curve). The other parameters, which are kept fixed, are $\mu=0.7$ and $\alpha=0.05$.}
 \label{Fig4}
\end{figure}
\begin{figure}[h!]
  \centering
  \subfigure[]{\includegraphics[width=0.48\textwidth]{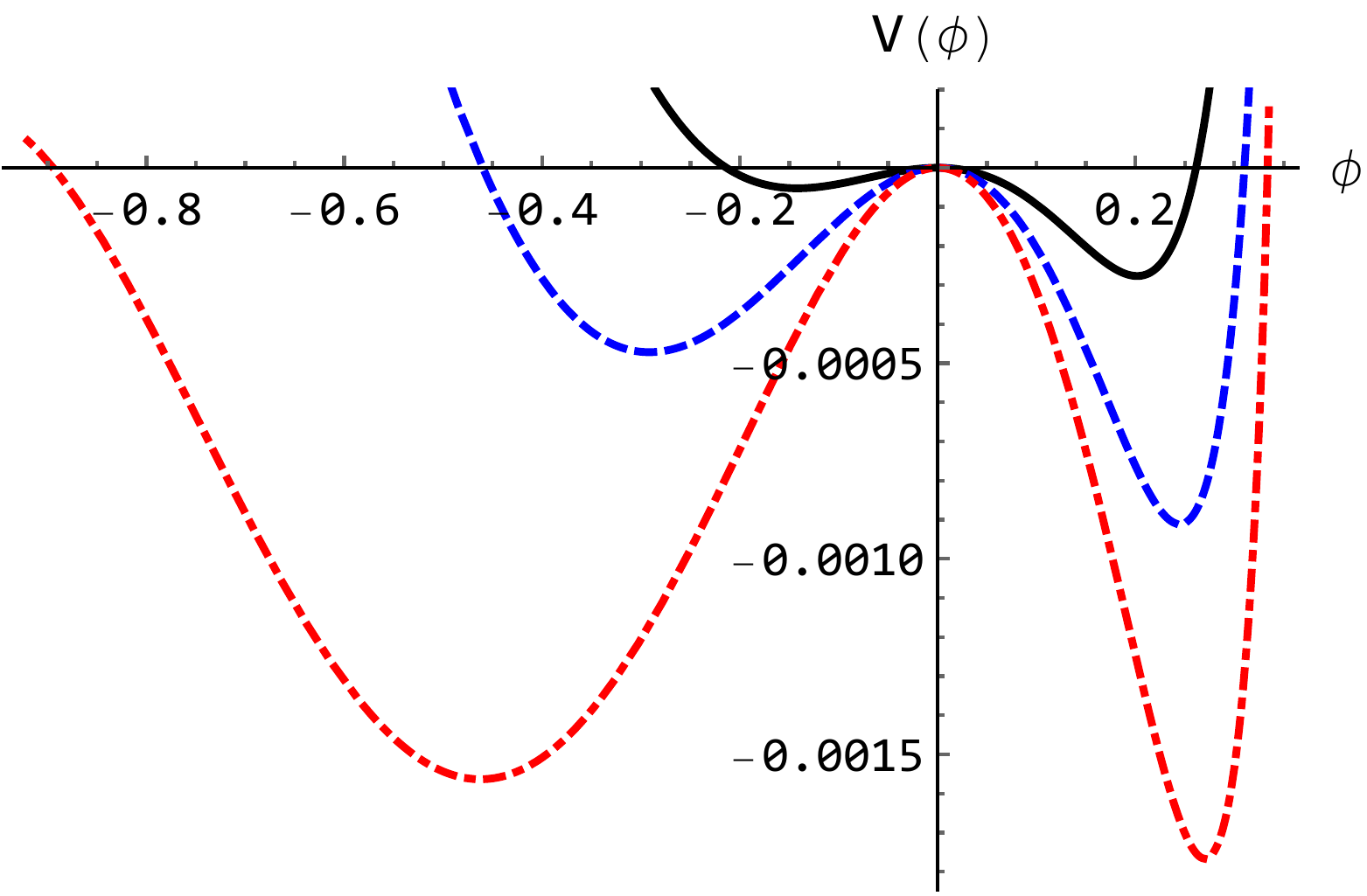}}
  \subfigure[]{\includegraphics[width=0.48\textwidth]{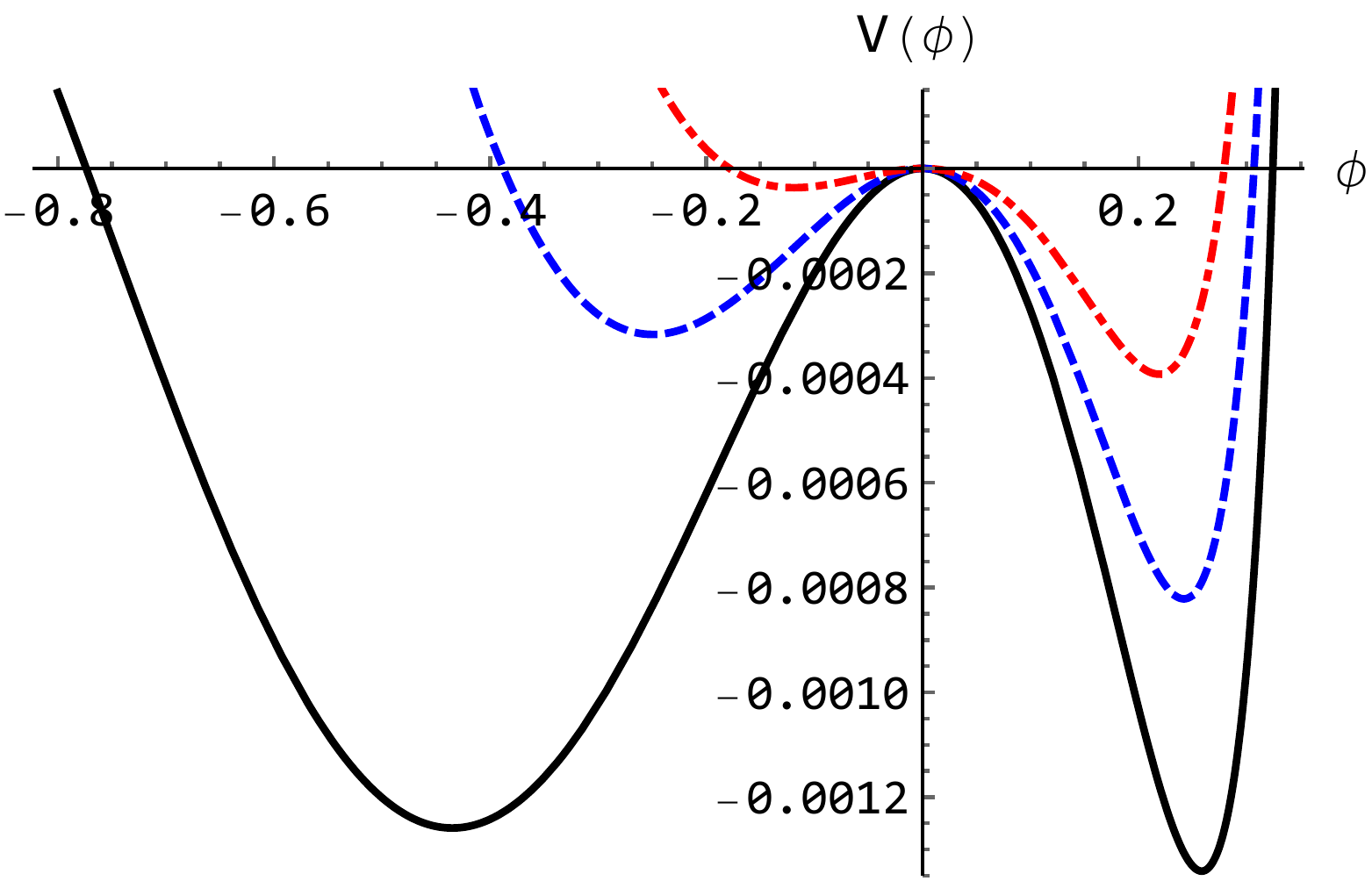}}
  \caption{The formation of the potential wells representing the coexistence of supersonic SWs
with $\phi>0$ and  $\phi<0$  (a) for $\alpha =0.266$, $\mu =0.3$ (solid curve), $\mu =0.35$ (dashed curve) and $\mu =0.4$ (dot-dashed curve);  (b) for $\mu =0.357$, $\alpha=0.26$ (solid curve), $\alpha =0.27$ (dashed curve) and $\alpha =0.28$ (dot-dashed curve).
 The other parameters, which are kept fixed, are ${\cal M}=1.5934$ and $\sigma=0.2$.}
  \label{Fig5}
\end{figure}
\begin{figure}[h!]
    \centering
    \subfigure[]{\includegraphics[width=0.48\textwidth]{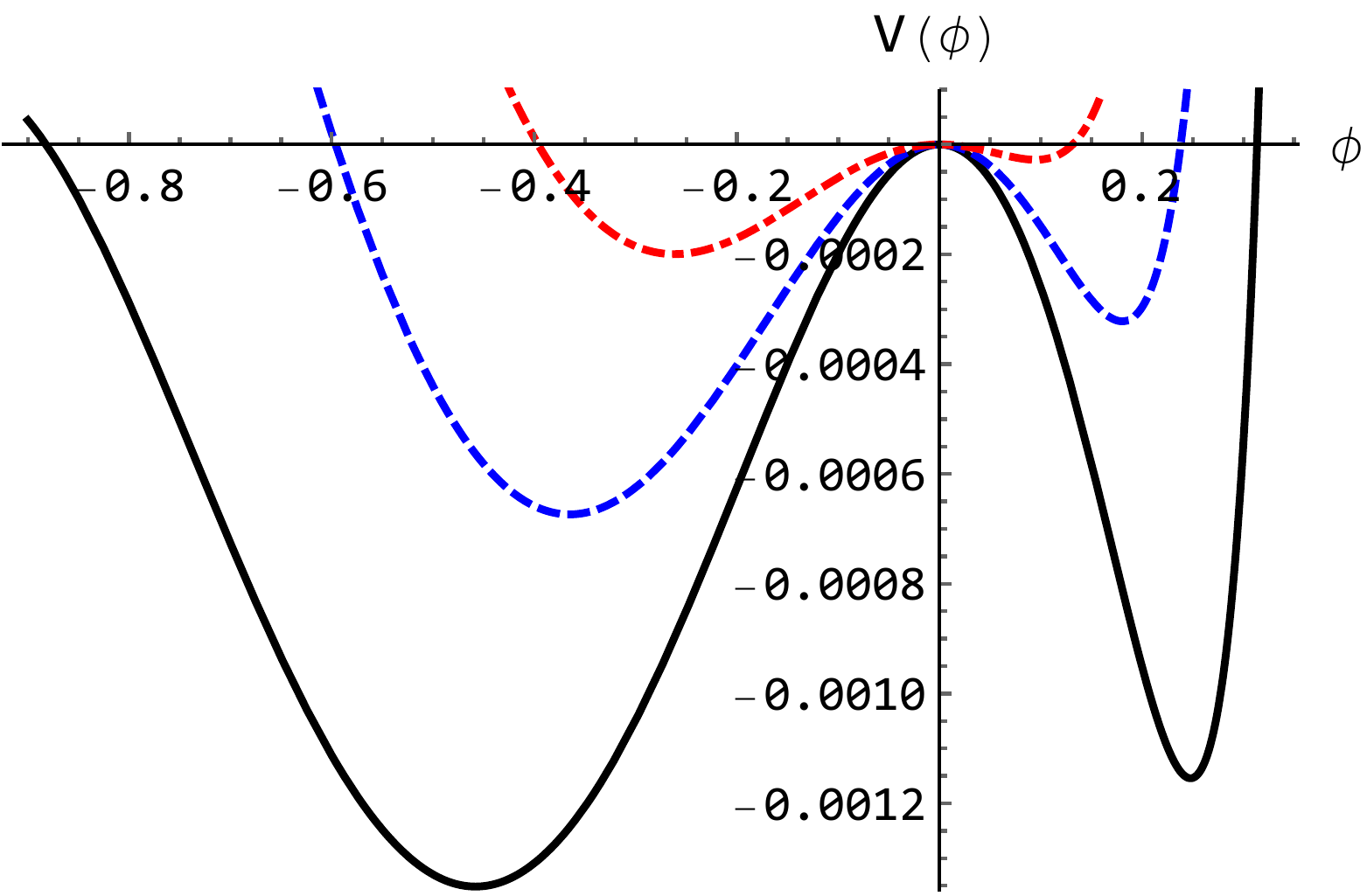}}
    \subfigure[]{\includegraphics[width=0.48\textwidth]{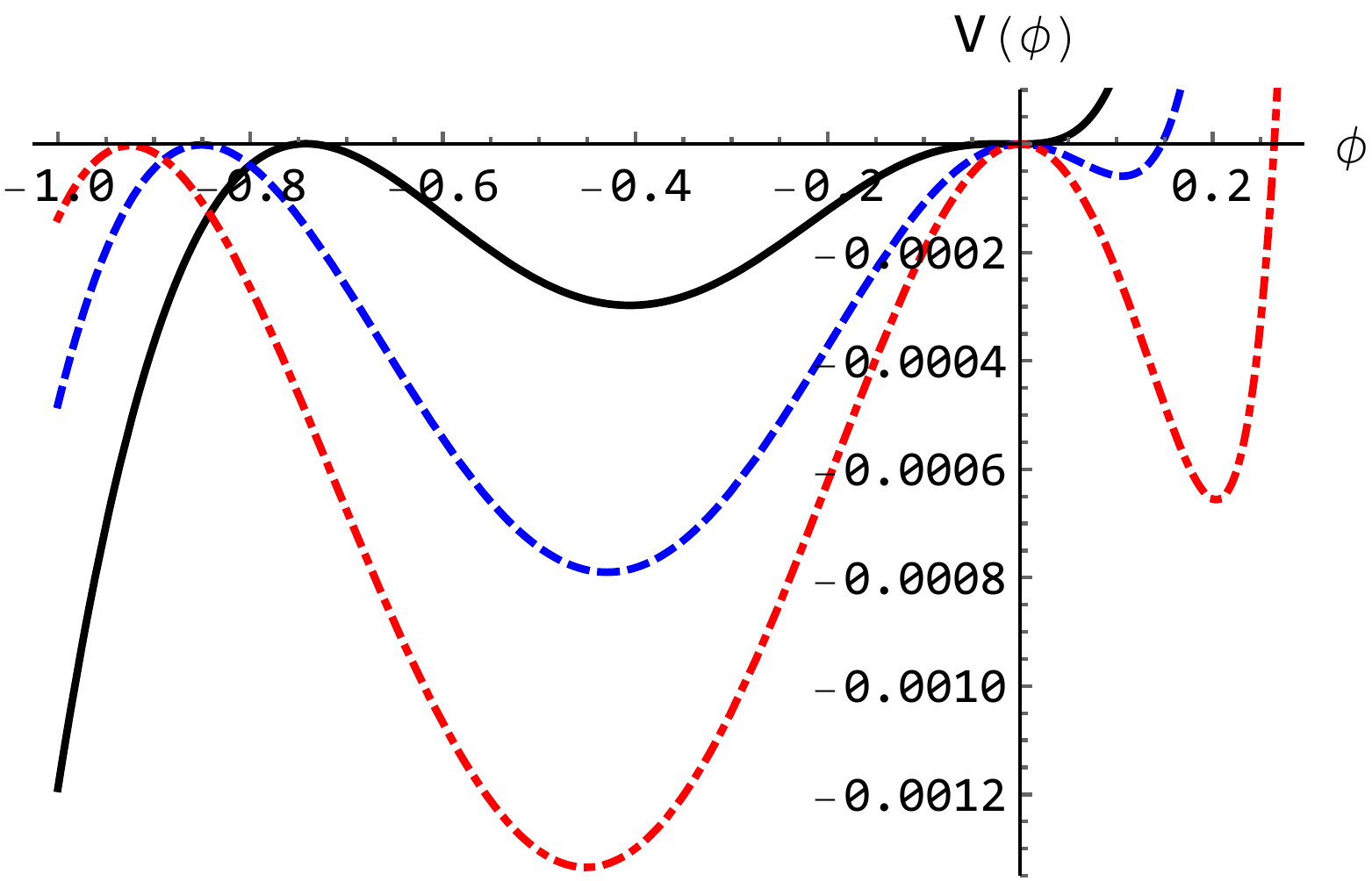}}
   \caption{The formation of the potential wells representing (a) the coexistence of supersonic SWs
  with $\phi>0$ and $\phi<0$ for $\alpha =0.258$, $\mu =0.36$, ${\cal M} =1.5868$, $\sigma = 0.2$ (solid curve), $\sigma = 0.22$ (dashed curve) and $\sigma = 0.24$ (dot-dashed curve); (b) the existence of  DLs with $\phi<0$ for ${\cal M}=1.4648$, $\alpha=0.25$ (solid curve), ${\cal M}=1.50362$, $\alpha =0.26$ (dashed curve), ${\cal M}=1.5452$, $\alpha =0.27$ (dot-dashed curve), $\mu=0.5$, and $\sigma=0.2$.}
 \label{Fig6}
\end{figure}
\begin{figure}[h!]
    \centering
    \subfigure[]{\includegraphics[width=0.48\textwidth]{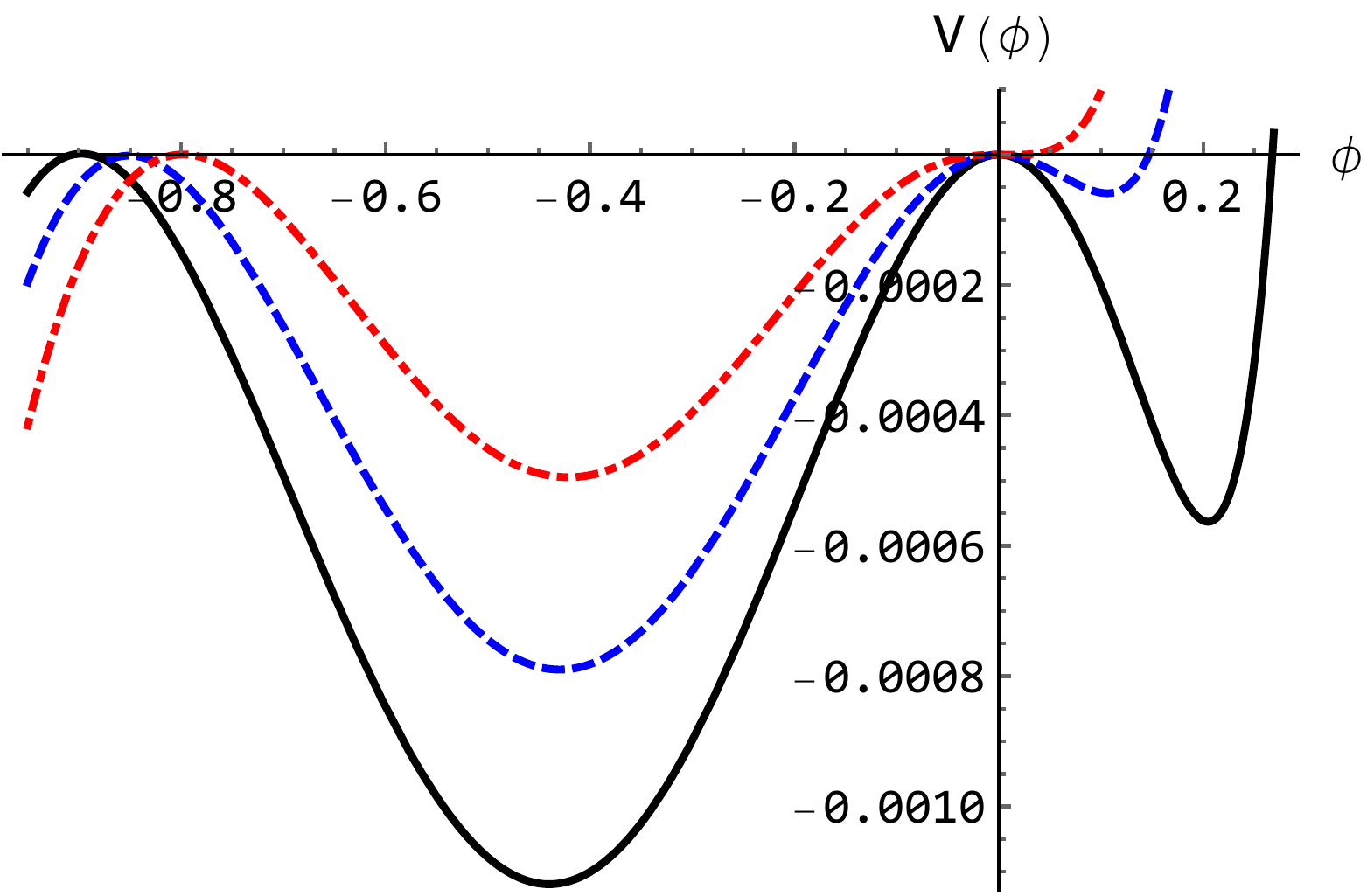}}
    \subfigure[]{\includegraphics[width=0.48\textwidth]{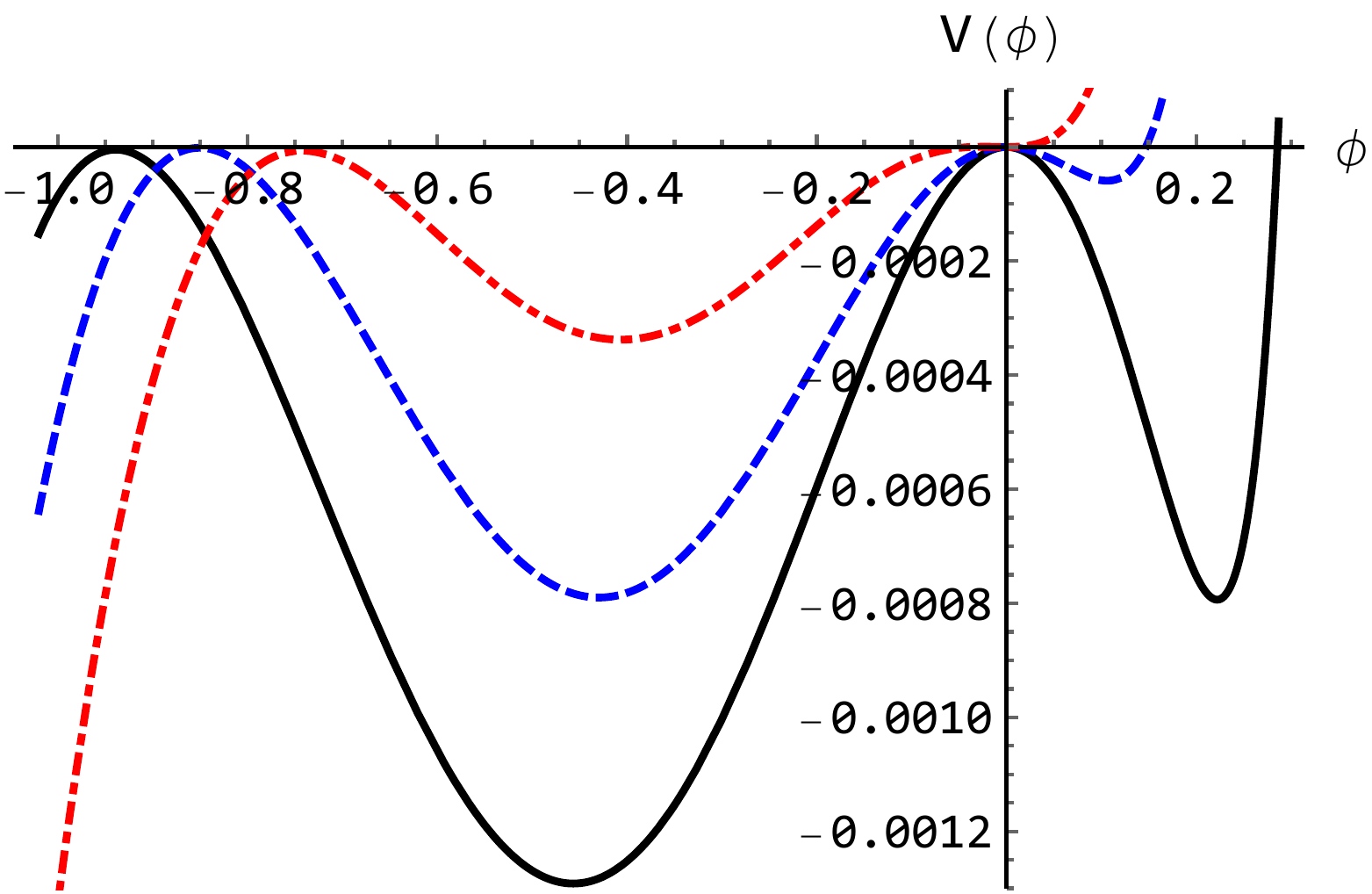}}
   \caption{The formation of the potential wells representing the DLs with $\phi<0$
(a) for ${\cal M}=1.4704$, $\sigma=0.15$ (solid curve), ${\cal M}=1.50362$, $\sigma=0.2$ (dashed curve), and ${\cal M}=1.5367$, $\sigma=0.25$ (dot-dashed curve), and $\mu=0.5$;
 (b) for ${\cal M}=1.5682$, $\mu=0.4$ (solid curve), ${\cal M}=1.50362$, $\mu=0.5$ (dashed curve), and ${\cal M}=1.4466$, $\mu=0.6$ (dot-dashed curve), and $\sigma=0.2$. The value of $\alpha = 0.26$ is kept fixed for both cases.}
    \label{Fig7}
\end{figure}
Figure \ref{Fig2}(b) shows how the parametric regimes for that existence of IA SWs with $\phi>0$ and for the coexistence of SWs with $\phi>0$ and $\phi<0$  changes with different plasma parameters. It means that the SWs with $\phi>0$ ($\phi<0$ and $\phi>0$ ) exist for the complex plasma parameters satisfying $V'''(0,\mathcal{M}_c)>0$ [$V'''(0,\mathcal{M}_c)<0$].  It is seen that the increase in the number density of the PCD species enhances the regime for the existence of the SWs  with $\phi>0$. The possibility for the formation of  SWs with $\phi<0$ as well as $\phi>0$ increases as the population of fast/energetic electrons increases. On the other hand, the rise of  the ion-temperature ($\sigma$) increases (decreases) the possibility for existence of SWs with  $\phi>0$ ($\phi<0$ as well as $\phi>0$).

Figures \ref{Fig3}-\ref{Fig7} can provide the visualization of the amplitude ($\phi_m$), which is the intercept on the positive or negative $\phi$-axis, and width ($\phi_m/\sqrt{|V_m|}$), where $|V_m|$ is the maximum value of $V(\phi)$ in the pseudo-potential wells formed in positive or negative $\phi$-axis. Figures \ref{Fig3} and \ref{Fig4} indicate the formation of the pseudo-potential wells in the positive $\phi$-axis, which corresponds to the formation of the subsonic IA SWs only with $\phi>0$, i.e. the subsonic IA SWs with $\phi<0$ does not exist
in the complex plasma system under consideration. The possibility for the formation of subsonic solitary wave increases (decreases) with increasing the value of $\mu$ ($\alpha$ and $\sigma$). It is seen that the amplitude (width) of the subsonic IA SWs decreases (increases) as we decrease the value of $\mu$. On the other hand, the amplitude (width) of subsonic SWs decreases (increases) with increasing the values of $\alpha$ and $\sigma$. It is worth to mention that the lower value of $\mu$ and higher values of $\alpha$ and $\sigma$ convert the subsonic SWs into supersonic ones. It is seen in figures \ref{Fig5} and \ref{Fig6}(a) that for $\mathcal{M} > \mathcal{M}_c$ the  supersonic SWs with $\phi>0$ and $\phi<0$ coexist. The amplitude (width) of supersonic SWs with both $\phi>0$ and $\phi<0$  increases (decreases) with increasing the value of $\mu$. On the other hand, the depth of potential wells (representing the coexistence of supersonic SWs with $\phi>0$ and $\phi<0$) decreases with increasing the values of $\sigma$ and $\alpha$.

The IA DLs only with negative potential is formed for $\mathcal{M} > \mathcal{M}_c$ as illustrated in figures
\ref{Fig6}(b) and \ref{Fig7}. The rise of the values of $\mu$ and $\sigma$ causes to decrease (increase) the amplitude (width) of the DLs (as shown in figure \ref{Fig7}). On the other hand, in figure \ref{Fig6}, the potential wells in the negative $\phi$-axis becomes wider as the nonthermal parameter $\alpha$ increases. It means that the amplitude of DLs are increased by the effect of nonthermal parameter, but the width of DLs decreases. It is noted here that for the formation of DLs, the increase in the values of $\alpha$ and $\sigma$ ($\mu$) is required a larger (smaller) value of the Mach number.
\section{Discussion}
\label{DIS}
We have considered a complex plasma medium containing Cairns nonthermally distributed electron species, adiabatically warm ion species, and PCD species, and  have investigated the IA SWs and DLs  in such a plasma medium.  We have employed the pseudo-potential approach which is valid for arbitrary or large-amplitude SWs and DLs. The results obtained from this theoretical work and  their applications can be  briefly discussed as follows:
\begin{itemize}
\item{The effect of the PCD causes to reduce the  IA wave phase speed, and to form subsonic SWs only with positive potential. On the other hand, the effects of Cairns nonthermal electron distribution and adiabatic ion-temperature cause to enhance the  IA wave phase speed, and to reduce possibility for the formation of the subsonic SWs, and finally convert the subsonic SWs into supersonic ones.}

\item{The amplitude (width) of the subsonic IA SWs increases (decreases) with the rise of the value $\mu$ and
${\cal M}$, but the amplitude (width) of the subsonic IA SWs decreases (increases) with the rise of the values
$\alpha$ and $\sigma$.  This is due to the fact that the phase speed of the IA waves decreases with rise of the value $\mu$,  but  increases with the rise of the value of $\alpha$ and  $\sigma$.}

\item{The supersonic  IA SWs with $\phi>0$ and $\phi<0$ coexist due to the presence of the certain amount of nonthermal or fast electrons (after a certain value of $\alpha$)  in the plasma system under consideration. However, the increase in the value of $\sigma$ and $\mu$ decreases the possibility for the formation of the IA SWs with
$\phi<0$.}

\item{The amplitude (width) of the  supersonic IA SWs (which coexist  with $\phi>0$ and $\phi<0$)  increases (decreases) as the values  of $\mu$ and ${\cal M}$ increase, but it decreases (increases) as the values  of
$\alpha$ and $\sigma$ increase.}

\item{The height (thickness) of the IA DLs (which exist only with $\phi<0$)  increases (decreases) as the values  of both parameters of its set  $\{{\cal M},~\alpha\}$  increase.  On the other hand, it decreases (increases) with the rise of the value of both parameters of their sets $\{{\cal M},~\mu\}$ and $\{{\cal M},~\sigma\}$.}
\end{itemize}

The advantage of the pseudo-potential method \cite{Cairns95,Mamun97,Bernstein57} is that it is valid for arbitrary amplitude SWs and DLs, but it does not allow us to observe the time evolution of the SWs or DLs.
To overcome these limitations, one has to develop a numerical code to solve the basic equations (\ref{IA-b1})$-$(\ref{ne}) numerically. This type of simulation  will be able to show the time evolution of arbitrary amplitude SWs and DLs. This is, of  course, a challenging research problem of recent interest, but beyond the scope of our present work.

To conclude,  we hope that the results of our present investigation should also be useful in understanding the basic features of the IA waves and associated nonlinear structures like SWs and  DLs in space environments (viz. Earth's mesosphere or ionosphere \cite{Havnes96,Gelinas98,Mendis04}, cometary tails \cite{Horanyi96}, Jupiter's surroundings \cite{Horanyi93,Tsintikidis96} and magnetosphere \cite{Horanyi93}, etc.) and laboratory devices \cite{Nakamura2001,Khrapak01,Fortov03,Davletov18}.
\vspace{-5mm}
\section*{Data availability}
Data sharing is not applicable to this article as no new data were created or analyzed in this study.
\vspace{-5mm}
\section*{Disclosure statement}
The authors declare that there is no conflict of interest.
\vspace{-5mm}
\section*{Acknowledgement}
A. Mannan gratefully acknowledges the financial support of the Alexander von Humboldt Stiftung (Bonn, Germany) through its post-doctoral research fellowship.
\bibliographystyle{tfnlm}
\bibliography{biblo}
\end{document}